\documentclass{ws-p9-75x6-50}

\begin{document}

\title{ Renormalization Group and  Dynamics of 
Supersymmetric Gauge Theories}

\author{K. Konishi}     

\address{Dipartimento di Fisica, Universit\`a di Pisa, \\
 INFN, Sezione di Pisa,  \\
Via Buonarroti, 2,  Ed. B  56127 Pisa, Italy
\\  E-mail: Konishi@df.unipi.it}


\maketitle

\abstracts{
We discuss questions   related to renormalization group and  to nonperturbative 
 aspects of non-Abelian gauge theories with  $N=2$ and/or   $N=1$ supersymmetry.   
 Results on perturbative and nonperturbative $\beta$  functions of these theories are reviewed, and new 
mechanisms of confinement and dynamical symmetry breaking recently   found in a class of $SU(n_c)$,  $USp(2n_c)$
and  $SO(n_c)$ theories are discussed. }

\section{Introduction }

We  start with a brief review on the NSVZ  $\beta$ functions in $N=1$
supersymmetric gauge theories and   related issues.

\subsection{NSVZ  $\beta $ function  in   $N=1 $ supersymmetric gauge
theories}\label{subsec:prod} The bare Lagrangian of an  $N=1$ supersymmetric gauge theory with generic
matter content is given by
\begin{equation}  L=  {1 \over 4} \int  d^2 \theta
\left( {1 \over g^2(M) }
\right) W^a W^a + h.c. +  \int d^4\theta \sum_i  \Phi_i^{\dagger}
e^{2V_i}  \Phi_i
\end{equation}
where
\begin{equation}   {1 \over g^2(M)}  = {1 \over g^2(M)} +  i\, {\theta(M)\over
8\pi^2}\equiv   i  {\tau(M) \over  4 \pi}
\label{holomorphy}
\end{equation}
and $g(M)$ and $\theta(M)$  stand for the bare   coupling constant and
vacuum parameter,   $M$ being the ultraviolet cutoff.   Note that with this convention 
the vector fields   $A_{\mu}(x)$ and the gaugino (gluino) fields  $\lambda_{\alpha}(x)$  contain the 
coupling constant and hence, in accordance  with  the non Abelian gauge symmetry,  are  not renormalized. 

By  a   generalized   nonrenormalization theorem \cite{NSV}
the effective Lagrangian at scale $\mu$ takes  the form,
\begin{equation}
L=  {1 \over 4} \int  d^2 \theta
\left( {1 \over g^2(M) } + {b_0 \over 8 \pi^2} \log {M \over \mu}
\right) W^a W^a + h.c. +  \int d^4\theta \sum_i Z_i(\mu, M) \Phi_i^{\dagger}
e^{2V_i}  \Phi_i \, ,
\label{eflag1}
\end{equation}
(plus higher dimensional  terms). Here
\begin{equation}
b_0= - 3 N_c  + \sum_i T_{Fi}; \qquad T_{Fi}= {1 \over 2}  \qquad ({\rm quarks})\, .
\end{equation}
Novikov et. al.   then  invoked    the 1PI  effective action to define a
``physical'' coupling constant
for which they obtained the well-known  $\beta$ function (Eq.(\ref{beta})
below) \cite{betafn}.
    
Recently the derivation  of the NVSZ beta function was  somewhat  
streamlined  by
Arkani-Hamed and Murayama \cite{AHM1,AHM2}.  (See also \cite{AFY}.)
 They obtained   the NVSZ beta function     in the 
standard Wilsonian framework,   without appealing to the 1PI  effective action
 (hence no subtleties due to zero momentum
external lines, such as
those leading to apparent violation of nonrenormalization
theorem\cite{NSV,Jones}).   They insist  
simply that at each infrared cutoff $\mu$  the matter kinetic terms  be
re-normalized so that it resumes the standard canonical form,  which is the standard
procedure in  the Wilsonian  renormalization group. 
But  the  field rescaling 
\begin{equation}
\Phi_i = Z_i^{-1/2} \Phi_i^{(R)} \, ,
\label{renphi}\end{equation}
introduces necessarily  anomalous  functional Jacobian \cite{KS},  
and one gets
\begin{eqnarray}
&L &=  \!\! {1 \over 4} \int  d^2 \theta
\left( {1 \over g^2(M) } + {b_0 \over 8 \pi^2} \log {M \over \mu}
 - \sum_i {T_F \over 8\pi^2} \log Z_i(\mu, M) \right) W^a W^a + h.c.   \\
&+& \!\!\int \! d^4\theta \sum_i  \Phi_i^{(R) \dagger} e^{2V_i}  \Phi_i^{(R)}
     \equiv    {1 \over 4 g^2\!(\mu)}\! \int  \!d^2 \theta
 W^a W^a + h.c. \!+ \!\int \!
d^4\theta \sum_i  \Phi_i^{(R) \dagger} e^{2V_i}  \Phi_i^{(R)} \, .   \nonumber
\label{effac}  \end{eqnarray}
where
\begin{equation}
{ 1\over g^2(\mu)} \equiv
{1 \over g^2(M) } + {b_0 \over 8 \pi^2} \log {M \over \mu}
 - \sum_i {T_{Fi} \over 8\pi^2} \log Z_i(\mu, M) \, .
\label{ghmu}
\end{equation}
This leads to the beta function (call it $\beta_h$ to distinguish
it from the more commonly used definition):
\begin{equation}
 {\beta_h}(g) \equiv  \mu {d \over d \mu}  \, g = - { g^3 \over 16
\pi^2} \left(
3 N_c -  \sum_i T_{Fi} (1- \gamma_i) \right) \, ,
\label{tildebeta}
\end{equation}
where
\begin{equation}
\gamma_i(g(\mu) ) =   - \mu { \partial \over \partial \mu } \log Z_i(\mu, M)|_{M, g(M)},  
\end{equation}
is the anomalous dimension of the $i-$th matter  field.
The same result follows by differentiating (\ref{ghmu}) with respect to $M$
with $\mu$ and $ g(\mu) $ fixed. 
 For SQCD these read 
 \begin{equation}
 {\beta_h}(g) = - { g^3 \over 16
\pi^2}  \left(  3 N_c -    N_f  (1- \gamma)  \right)     \, ,
\qquad 
\gamma(g) =  
 - {g^2 \over  8 \pi^2}  {N_c^2 -1 \over  N_c}  + O(g^4),  
\label{tildebeta1}   \end{equation} 
Eq.(\ref{tildebeta}) and Eq.(\ref{tildebeta1})   are the NSVZ  $\beta$  functions \cite{betafn}. 

Note that  the ``holomorphic'' coupling constant   $g(\mu) $    is a perfectly good
definition
of the effective coupling constant:  it is finite as
$   M \to \infty; \quad \mu = {\hbox{\rm finite}}, $
and  physics  below $\mu$ can be computed in terms of it.
Vice versa,     the coupling constant  defined as the inverse of the coefficient of $W^a W^a$  in (\ref{eflag1}),
$({1 \over g^2(M) } + {b_0 \over 8 \pi^2} \log {M \over \mu})^{-1}$,     is {\it
not} a good
definition of an effective coupling constant,  as long as $N_f \ne 0$:  it
is divergent in the
 limit the ultraviolet cutoff is taken to infinity.
In other words,  the renormalization of the matter fields (\ref{renphi}) is the standard,  compulsary  step  of 
renormalization, such that the low energy physics is independent of  the ultraviolet cutoff, $M$. 

Let us also   note that,  in spite of its  name, the holomorphic coupling constant
gets renormalized in a non-holomorphic way, due to the fact that $Z_i(\mu, M)$
is real.
Another consequence of the reality of $Z_i(\mu, M)$ is that   $\theta$  is not renormalized:   this is evident from the same RG equation
(\ref{tildebeta}) written in terms of $\tau$ variable ,
\begin{equation}    \mu {d \over d \mu}  \, \tau(\mu)   = - { i \over 2 \pi}  \left(
3 N_c -  \sum_i T_{Fi} (1- \gamma_i) \right),  
\end{equation}
showing that   the NSVZ beta function is essentially perturbative. 
It is interesting to observe that    the above procedure  parallels  nicely the original derivation by Novikov et. al.
of  the beta function by use of some  instanton-induced  correlation functions.  
More recently  the NSVZ  $\beta$ function in $N=1$  supersymmetric QCD has been  rederived 
by  Arnone, Fusi  and Yoshida \cite{AFY},  by using
the method of exact renormalization  group.  

\subsection{Zero  of the NVSZ beta function and Seiberg's duality and  CFT   in
SQCD}\label{subsec:wpp} 
For  the range of the flavor  $ {3N_c \over 2} < N_f < 3 N_c  $    (conformal window)  Seiberg discovered by using the 
NVSZ $\beta$   function  that 
the theory  at low-energy is at a nontrivial   infrared fixed point \cite{Sei}.     At the zero of the $\beta$   function 
the anomalous dimension of the matter field is  found to be:
\begin{equation}    \gamma(g^*)  = {3 N_c -    N_f  \over  N_f}. 
\end{equation}
It turns out that this result is in agreement with 
that  determined  from the superconformal algebra, which contains the non-anomalous   $U_R(1)$ symmetry. 
This and many other consistency checks allowed Seiberg to conclude that   in the conformal window,
and at the origin of  the moduli space (namely, in the theory  where all VEV's vanish),    the theory has a nontrivial 
infrared  fixed point.  Such a theory has no particle description,  and as such, can be described by more than 
one type of  gauge theory.    In fact, in $SU(N_c)$ theory,  the theory can be either described as the standard   
SQCD with $N_f$ flavors, or  in terms of a dual theory, which is  an $SU({\tilde N}_c)$ gauge theory   with   $N_f$  sets of dual
quarks, plus  singlet meson fileds, where ${\tilde N}_c \equiv   N_f-N_c$.  They have  the same  infrared  behavior.   This is
the first  example of the  $N=1$ non-Abelian  duality,   found in many other  theories  subsequently.

This development enabled Seiberg   to  complete  the picture  of  
  dynamical properties of $N=1$  supersymetric  QCD  in  all cases.     Phase,  the low-energy effective degrees
of freedom, effective gauge group,   etc. are summarized in  Table 1  (where the bare quark masses are taken to be zero). 

\begin{table}[h]      
  \caption { Phases of $N=1$ supersymmetric  $SU(N_c)$ gauge theory with $N_f$ flavors.    $M_{ij} ={\tilde Q}_i  Q_j  $
and  $B= \epsilon_{a_1 a_2 \ldots a_{N_c } }  \epsilon^{i_1  i_2 \ldots  i_{N_c}   }   Q_{i_1}^{a_1}\,  Q_{i_2}^{a_2} \dots
Q_{i_{N_c}}^{a_{N_c}} $
(${\tilde B}$  is constructed similarly  from the antiquarks, ${\tilde Q}$'s)  stand for the meson and baryon like 
supermultiplets.    
"Unbroken" means that the full chiral symmetry  $G_F=SU_L(N_f)\times SU_R(N_f)\times U(1)$
is realized linearly at low energies. 
Actually,   for $N_f >  N_c$   continuous vacuum degeneracy of the theory  survives quantum  effects, and the
entities in the table  refers to a  representative vacuum at the origin of the quantum moduli space (QMS).
For the special case of   $N_f=N_c$, the QMS  is parametrized by  
$\det M -  B {\tilde B} =\Lambda^{2N_f}$ and the  $U(N_f)$ symmetry  is  the unbroken symmetry of the vaccum
with      $M= {\mathbf 1}  \cdot \Lambda^{2},\,\, B=0$.            }
\begin{center}
\footnotesize
\vskip .3cm
\begin{tabular}{|ccccc|}
\hline
 $N_f$    &   Deg.Freed.      &  Eff. Gauge  Group
&   Phase    &  Symmetry     \\
\hline
\hline
$0$ (SYM)   &   -    &   -              &   Confinement
   &     -           \\ \hline
$ 1 \le N_f  <    N_c $            &  -     & -       &
  no vacua        &    -      \\ \hline
$N_c$    &  $M, B, {\tilde B} $        &   -
  &    Confinement
&          $U(N_f)$ 
\\ \hline
$N_c +1$   &   $ M, B,{\tilde B} $       &   -
  &    Confinement
&         Unbroken   
\\ \hline
$N_c+1 < N_f < {3N_c \over 2} $    &   $ q, {\tilde q}, M  $     &   $ SU({\tilde N}_c ) $     &   Free-magnetic
&        Unbroken
\\ \hline
$ {3N_c \over 2} < N_f < 3 N_c  $   &   $q, {\tilde q},M  $   or   $Q, {\tilde Q} $    &   $SU({\tilde N}_c ) $  or  $ SU(N_c) $   &  SCFT
&        Unbroken  \\ \hline
$N_f  = 3N_c$   &   $Q, {\tilde Q} $   &   $SU(N_c ) $   &   SCFT (finite)
&        Unbroken  
\\ \hline
$N_f > 3N_c  $  &  $Q, {\tilde Q} $   &
$ SU(N_c)  $                &  Free Electric  
&    Unbroken     \\ \hline
\end{tabular}
\label{tabsun}
\end{center}
\end{table}

\subsection{Meaning of the pole of the NVSZ beta function}

The status of the denominator of the so-called  NSVZ  $\beta$ function is subtler. 
Although  it is not  necessay, one might wish   to make a further {\it finite}     renormalization 
in Eq.(\ref{effac})    
to get   the canonical form of gauge kinetic terms,
$ -F_{\mu \nu} F^{\mu \nu}/4  +   i  \lambda  \sigma_{\mu}  D^{\mu}  {\bar  \lambda}. $  
The redefinition needed is
\begin{equation}
A_{\mu}= g_c A_{c \mu}, \quad      \lambda=g_c \lambda_c.  \label{chvar}\end{equation}
 This introduces as functional--integral
Jacobian   an extra factor \cite{AHM2},
\begin{equation}
\exp  \, {1 \over 4} \int d^4x  \int d^2\theta   \,  {N_c  \log g_c^2
\over 8 \pi^2 } \,  \, W^a W^a + {\rm h.c.}
\end{equation}
and as a consequence,   leads to  the change of the coupling constant
\begin{equation}
{1 \over g^2} = {1 \over g_c^2} + { N_c  \over 8 \pi^2 } \log g_c^2 ,
\label{gcanon}
\end{equation}
introduced    earlier  by   Shifman and  Vainshtein \cite{SV91}  as a way  to reconcile holomorphy and 
renormalization. 
The well-known NSVZ beta function  of the form \cite{betafn} 
\begin{equation}
\beta(g_c)=- { g_c^3 \over 16 \pi^2}
{ 3 N_c -  \sum_i T_{Fi} (1- \gamma_i) \over
1- N_c g_c^2/8\pi^2} \, .
\label{beta}
\end{equation}
follows then from (\ref{gcanon}) and
(\ref{tildebeta}).
   In the case of $N=1$
   pure   Yang-Mills theory
such a pole of the beta function     has led to an interesting 
conjecture by Kogan and Shifman    that   there is another, dual phase of the theoy in which  
 the coupling
constant $g_c$ grows in the ultraviolet \cite{denombeta}.

  However,  the very way   the denominator   arises   from the functional change of variables, 
  (\ref{chvar}), (\ref{gcanon}),   reveals the meaning of such a behavior of the beta function. 
 In fact, the right hand side of
(\ref{gcanon}) has a minimum at
$g_c^2= 8\pi^2/N_c$,  precisely corresponding to  the pole of the NSVZ beta
function,   where it takes the
value,
\begin{equation}
{N_c \over 8 \pi^2 } \log{8 \pi^2 e \over N_c}    > 0
\end{equation}
(for    $N_c   <   215$).
On the contrary,    the left hand side of
(\ref{gcanon}) evolves down to zero if the beta function has no zero
($N_f < 3 N_c /2$).   Thus
for large values of $g$ ($g  >    8\pi^2/N_c \log(8 \pi^2 e/N_c)$)
the redefinition
(\ref{gcanon}),   with a real ``canonical coupling constant'',   is not
possible.    In other words,    the change of the functional variables involved is not a 
proper one,   the new variable    $A_{c \mu} $  being  complex. 

The runnings of the holomorphic and canonical coupling constants are compared  in Fig. 1    for the case of 
$N_f=0$,   from which one  sees  that   both the absence of evolution below the scale $ ({8 \pi^2 e
/ N_c})^{1/3}  \Lambda $
 and  the apparent new phase of the theory  are    artifacts
 caused by   the improper  change of the variable (\ref{gcanon}).
The  pole of  the beta function     signals   the failure of
$g_c$ as a coupling constant (and $A_{c \, \mu }(x)$  as a functional
variable).

\begin{figure}[t]
\begin{center}  
\epsfxsize=15pc 
\epsfbox{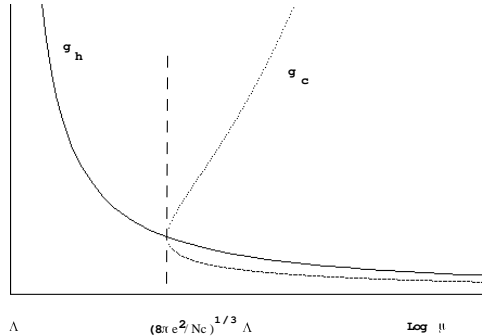} 
\end{center}
\caption{Evolution of the holomorphic coupling constant versus that of the "canonical" coupling constant. }  
\end{figure}

This however means that if one starts in the UV    by using the
standard ``canonical'' coupling constant
and  studies the  RG evolution towards the low energies,   one must switch
to the ``holomorphic''
description  at certain point (in any case,   before the ``critical'' value
$\alpha_c= 2\pi  /N_c$ is reached),
   in
order to describe  physics smoothly down to $\mu = \Lambda.$
 The impossibility of writing  a low energy effective Lagrangian with
canonically normalized
gauge kinetic terms,  does not represent   any inconsistency, since   
the low-energy  physical degrees of
 freedom   are    not     gauge (and quark)  fields themselves.
This last  statement does not apply for   $N_f >  3  N_c$,  but  there is no obstruction in
using the canonical  variables in these cases, since the theory is infrared-free.

 The success of the NSVZ beta function in the case  of SQCD
in the conformal window ($3 N_c/2  < N_f < 3 N_c$)   discussed already,    
especially the
determination of the anomalous dimension of the matter fields
at the infrared fixed point,
  does not    require the use of the canonical
coupling constant.     This is important because
the  anomalous dimensions at the IR fixed point are  physically significant 
numbers.

\subsection{Exactness of the NVSZ beta function}

One might wonder how ``exact'' all this  is.  It is clear that the
diagrammatic  proof of
the  generalized  nonrenormalization theorem of \cite{NSV}  used in Eq.(\ref{eflag1})   is  valid only
within perturbation
theory.

It was argued  on the other hand in \cite {AHM2}  that due to the existence
of an anomalous $U_R(1) $ symmetry
 the   beta  functions are purely
perturbative,  hence   the NSVZ beta function is exact perturbatively and
nonperturbatively,
at least for pure $N=1$  Yang--Mills theory.
 In fact,   the (holomorphic)   coupling constant at scale $\mu $  must
satisfy
\begin{equation}    \tau(\mu) =    \tau(M) +   f\left(\tau(M),     \mu/M\right) \, ,
\end{equation}
where  
$f$ is a holomorphic function of $\tau$.
It follows that 
$\beta(\tau) =  \mu (1/\mu) \tau(\mu)$  
shares the same property.    Together with the
periodicity in $\theta$
with period $2 \pi$, one finds that
\begin{equation}  
\beta(\tau) =
 \sum_{n=0}^{\infty}  a_n  e^{2 \pi i  n \tau   },
\end{equation}
where $a_n$ is the $n$- instanton contribution.   If   the right hand
side  is independent of $\theta$  it must consist    only of  the
perturbative term, $n=0$.  That leads back  to the  NSVZ $\beta$ function. 

This   argument is however only valid  in theories
in which  the anomalous  $U_R(1)$   
symmetry   is not spontaneously broken by the VEVS of some scalar field.
Examples are  the pure $N=1$ Yang--Mills theory or  the    $N=1$ SQCD at the
origin of the space of
vacua   (with all scalar vevs vanishing):   there   the argument of
\cite{AHM2}   is valid and
the NSVZ  $\beta$ function is exact perturbatively and nonperturbatively.

Vive versa,    in  a generic   point of the space of vacua
 of $N=1$     SQCD,  or   at any point  of  a   $N=2$   supersymmetric
Yang--Mills theory (a $N=1$ supersymmetric gauge  theory    with a matter
chiral multiplet   in the adjoint representation),   for example,
$U_R(1)$  invariance is spontaneously broken \cite{Konishi},  i.e.,  by anomaly as well as by the 
VEVS  of certain gauge invariant composite fields.   There is a
nontrivial $\theta$ dependence.    By holomorphic dependence of $\beta$ on
$\tau= {\theta \over
2\pi } + {4\pi i\over  g^2}$  this implies that the beta function gets
necessarily   instanton corrections.  

\subsection{Other developments}

One interesting  development involves the existence in many models of infrared fixed lines,
as can be shown by using the explicit formula for the supersymmetric anomaly multiplet \cite{IRfixed}.    
In other words in these models there are exactly marginal operators. 
Another  very interesting  development deals with the possible $c-$ (or $a-$) theorem in four dimensional
supersymmetric  gauge theories \cite{anselmi}.        There remain  some longstanding problems such as the "${4 \over 5}$
puzzle"   in the  computation of the gaugino condensate in the super Yang Mills theory, which is made     more accute
after  some recent analysis \cite{4over5}.

\section {$N=2$ gauge theories  }
\subsection{ $\beta$ function} 
The (bare) Lagrangian of the  pure    $N=2\, $     $SU(2)$   Yang--Mills theory reads
 \begin{equation}   {1\over 4\pi} \hbox{\rm Im}    \, \tau_{cl} \left[\int d^4 \theta \,
\Phi^{\dagger} e^V \Phi +\int d^2 \theta\,{1\over 2} W W\right], 
\end{equation}
\begin{equation} \tau_{cl}={\theta_0 \over 2\pi} + {4 \pi i \over g_0^2}, \label{struc}\end{equation}
where 
$\Phi= \phi + \sqrt2 \theta \psi +\ldots,$ and $W_{\alpha} = -i \lambda + {i \over 2}
(\sigma^{\mu} {\bar{\sigma}}^{\nu})_{\alpha}^{\beta}  F_{\mu \nu} \theta_{\beta} +\ldots $ are both
in the adjoint representation of the gauge group.  
$N=2$ supersymmetry   restricts the form of the  low-energy effective action to be 
\begin{equation}  L_{eff} = {1\over 4\pi} Im \, [ \int d^4\theta \, {\partial F(A) \over \partial A} {\bar
A} +  \int d^2 \theta {1\over 2} {\partial^2 F(A) \over \partial A^2} WW ], \end{equation}
where $F(A)$, holomorphic in $A$,  is called prepotential. As seen from this
expression (the VEV of ) $ {\partial^2 F(A) \over \partial A^2} $ plays the role of the
effective $\theta$ parameter  and coupling constant, 
$\tau(a)= {\partial^2 F(a) \over \partial a^2}=\theta_{eff}/2\pi+4\pi i/g_{eff}^2.$
Perturbative and nonperturbative (instanton) corrections lead to the general form
\begin{equation}  F(a)={i \over 2\pi}A^2 \log {A^2\over \Lambda^2} + F^{inst},
\qquad F^{inst}=\sum_{k=1}^{\infty}     c_k    (\Lambda/A)^{4k} A^2,\end{equation}
is the contribution from
multiinstanton effects.    The prepotential $F$ has been found  by Seiberg and Witten \cite{SW}
through   the introduction  of an auxiliary   curve (torus),   which in the simplest case of the pure $SU(2)$ $N=2$ Yang-Mills
theory reads,
\begin{equation}     y^2 =   (x^2 - \Lambda^4) (x-u), \qquad  u= \langle   \hbox{\rm Tr}   \Phi^2  \rangle ,    \end{equation}
and    $d a_D / du$ and   $da / du$  are  given by  the period     integrals 
\begin{equation}   d a_D / du= \hbox{\rm const}  \oint_{\alpha} {dx \over y}; \quad d a / du= \hbox{\rm const}  \oint_{\beta} {dx \over y}.  \end{equation}
Such a construction has then been generalized to  $SU(n_c), USp(2n_c)$ and  $SO(n_c)$  gauge groups  with an
arbitrary number  of flavors \cite{others}.

Due to the holomorphic
nature of Wilsonian effective action  the RG equation can be cast
into the form \cite{Seiberg2}      
\begin{equation}
\beta(\tau)\equiv   \mu { d \tau\over d \mu}  = { 2 i \over \pi} \,( 1 +
 c_1 \, e^{2 \pi i  \tau} + c_2 \, e^{4 \pi i  \tau}+ \ldots)
\label{RG1}
\end{equation}
for $\,  \hbox{\rm Im} \,  \tau \gg 1\,$   (or $\,g^2 \ll 1\,$)  where
\begin{equation}
\tau= { \theta \over 2 \pi} + { 4 \pi i \over g^2} \, ,
\end{equation}
and  $\mu$ is the scale.  Recently  several papers   discussed 
the calculation of  the  exact, nonperturbative $\beta$  function.\cite{n2betafns} 
All of these  ``$\beta$  functions"    have the correct UV behavior
by construction, but none of them has the correct IR  behavior.   
The correct behavior at $\tau \sim 0$ can be found by noting that    in the theory near 
$u=\Lambda^2$  the IR cutoff is given by  $\mu_{IR}= \sqrt2 |a_D|$.
The leading behavior is that of a dual  QED with a single monopole,  and transformed   back to 
the electric description by a duality transformation it gives 
\begin{equation}
\beta(\tau)   \sim {1 \over  i \pi \tau_D^2}  =  -   {i \over  \pi}  \tau^2
\quad   \quad \rm{ as }
\quad \tau \to 0,
\label{truebeta}
\end{equation}
For  CP invariant cases ($\theta=0$)  this means the behavior
\begin{equation}  \beta(g)\sim    -  {2 \over g},\end{equation}
at large $g$.  See Fig. 2.   
\begin{figure}[t]
\begin{center}  
\epsfxsize=13pc 
\epsfbox{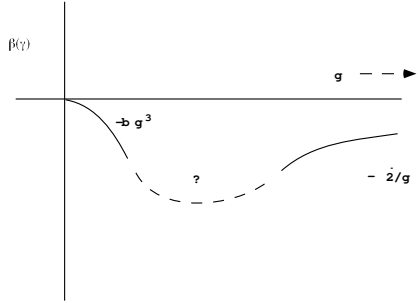} 
\end{center}
\label{betaog}
\caption{The $\beta$ function  of the theory for     $u$ real and   $u \ge   \Lambda^2$.}  
\end{figure}
Actually,    the "$\beta$ function"  computed by these authors    is equal to  
$   2  u   { d \tau_{eff} \over  d u}, $
where both  $\tau_{eff}$ and $ u$ are renormalization-group invariants.   It corresponds to  the variation   of the
low-energy  effective coupling constant {\it within}  the QMS, not to the standard $\beta$ function.

\subsection{Nonrenormalization of  $a$  and renormalization of $\theta$   in pure $N=2$ Yang-Mills theory}

Even though  the exact, nonperturbative $\beta$ function still eludes us,     the exact Seiber-Witten solution 
yields     a result  which amounts to the {\it integral }  of the renormalization group equation.  
Namely,     in  each theory  (i.e., at  each point of QMS) characterised by $u$,    both 
the bare  and  the corresponding low-energy  $\theta$  parameters are knowm  exactly.  The key relation is 
\begin{equation}    a= a^{(cl)}. 
\label{keyrelation} \end{equation} 
This relation has the following meaning:  on the left hand side, one has 
$a\equiv  \langle  A \rangle  $, where   $A$  is the low-energy scalar supermultiplet, $N=2$ superpartners of
the  photonlike  gauge multiplet  $W$.    The right hand side is the classical VEV of  the adjoint field, 
\begin{equation}       \langle   \phi  \rangle   = { 1\over 2}  \pmatrix{ a^{(cl)} & 0 \cr 0 & -a^{(cl)}}.  \label{exact}  \end{equation}
To show (\ref{keyrelation}),   note that 
because of the exact mass formula, 
\begin{equation}  M_{n_m, n_e}  = \sqrt2 \, | n_m a_D +  n_e a|, \end{equation}
$|a|$ represents the physical mass of a $(0,1)$  particle  ($"W^{\pm}"$ bosons or  their fermion partners).
Classically, it can be read off from  
the Lagrangian \begin{equation}   \sqrt 2   \, \,  \hbox{\rm Tr}  \, \phi^* [\lambda, \psi],     
\end{equation} 
and  is indeed equal to    $ \sqrt2  \,  |a^{(cl)}|.$     As  $A$  is a $N=2$ superpartner of the Yang-Mills field $A_{\mu}$
($gA_{\mu}$  in the canonical  definition) it is not renormalised perturbatively.   
In order to  see whether   $a$    gets  nontrivial  instanton contributions  one must compute, e.g.,
\begin{equation}   \langle  \lambda^+(x)  \psi^-(y)  - \psi^+(x) \lambda^-(y)  \rangle .
\label{wmass}  \end{equation}
This Green function obviously gets  the classical  contribution proportional to 
\begin{equation}  \sqrt2  \, a^{(cl)}   \int  d^4z  \, S_F(x-z)   S_F(y-z).    
\end{equation}
By studying the  possible instanton corrections to (\ref{wmass})  it can be shown that there are 
no instanton corrections whatsoever   to this result,   to any  instanton number.  The reason is that   because of the
symmetry  of the classical Lagrangian, there are  four "supersymmetry zero modes" (two for $\lambda$, two for $\psi$)
to any instanton number:
only two of them are  eliminated in the Green function (\ref{wmass}) hence the functional integration over 
the other zero modes yields  a vanishing  result.    Eq. (\ref{keyrelation}) is thus exact perturbatively and 
nonperturbatively.

This is to be contrasted with   the case of the one point function,  $ u\equiv  \langle  \hbox{\rm Tr}  \phi^2 \rangle $, or the Green function
\begin{equation}   \langle  \lambda (x)  \lambda (y)  \psi (z) \psi (w)  \rangle ,
\label{fourp}  \end{equation}
in which case  nonvanishing contributions are found  to all  instanton numbers.     

Let us note that the relation Eq. (\ref{keyrelation})  is implicitly assumed   in all  direct  instanton checks   of the 
Seiberg-Witten curves \cite{instantons}.  

An immediate consequence of   (\ref{keyrelation})  is that    the bare and renormalized  $\theta$ 
can be computed for each $u$:
\bea  \theta_{UV} &=& \theta_{bare}=    4 \, \hbox{\rm Arg}  \,  a^{(cl)}= 4 \, \hbox{\rm Arg}  a;  \nonumber \\
   \theta_{IR} &=& \theta_{eff}=  2 \pi   \, {\hbox{\rm Re}} \,  \tau_{eff} =  2 \pi  \, {\hbox{\rm Re}}  {da_D \over da}. 
\eea
 This instanton-induced renormalization effect is illustrated for several representative points of QMS in Table \ref{tabtheta}.
Thus in a generic point of QMS in $N=2$  supersymmetric pure Yang-Mills theory  $\theta$  grows in the infrared. 
This is opposite  to what was found in a model with soft supersymmetry breaking    where  CP
 violation was found to be    suppressed in the infrared  by the instanton effects \cite{KK}. 
\begin{table}[h]   
\caption{Instanton-induced renormalization of $\theta$  }
\begin{center}
\begin{tabular}{|c|c|c|c|}
\hline
 $ u/\Lambda^2   $   & $\theta_{UV} $
&  $\theta_{IR} $   &  Comment     \\
\hline
\hline
$2$             &       $0$       &      $0$          &        CP           \\ \hline
$-2+0.0001  i $  &  $\sim 2\pi $    &   $\sim 2\pi $      &      $\sim$ CP              \\ \hline
  $i$        & $\pi$             &   $\pi$    &         CP      \\ \hline
   $1+ 0.01 i  $       &  $0.3305$    &  $0.8554$    &       strong coupling     \\ \hline
 $1.1+ 0.01 i  $      & $0.2560$      &    $0.4968$    &        strong coupling    \\ \hline
 $0.9+ 0.01 i  $      & $0.4627$      &    $1.3310$    &        strong coupling    \\ \hline
 $0.5+ 0.01 i  $      & $1.1966$      &    $2.3789$    &            \\ \hline
  $4+ 4 i  $      & $1.5786$      &    $1.5903$    &      semi-classical   \\ \hline
\end{tabular}
\label{tabtheta}
\end{center}
\end{table}

\begin{figure}[h]
\begin{center}
    \epsfig{file=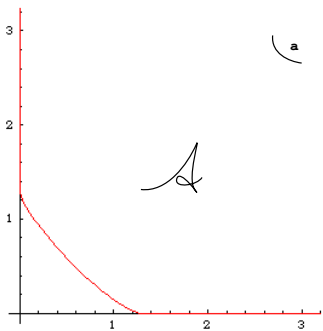,width=6cm}    
\end{center}
\caption{Fundamental region of the classical space of vacua (CMS) }
\label{fundaa}
\end{figure}

The key relation  Eq.(\ref{keyrelation})   furthermore allows to give a precise meaning to the statement that 
classical and quantum moduli  space are equal  in the  pure $N=2$ Yang-Mills theory.   Namely,  if the 
fundamental region   $a$  space is defined as the image ${\cal  A}$     of the  upper half plane of $u$  (Fig.\ref{fundaa}),    
 then the exact statement is 
 \begin{equation}    (CMS)_{\cal A}=QMS. 
\end{equation} 
Note that other regions of  CMS  ($a^{(cl)}$ outside  ${\cal A}$) do not represent inequivalent vacua: 
each such theory is equivalent to some of the theory in   ${\cal A}$  as they are related by an $SL(2, Z)$
transformation  (monodromy  transformations in QMS).

\subsection{The zero of the $\beta$ function  at  the infared fixed points (SCFT) of $N=2$ susy gauge theories}

A very interesting development was the discovery of the large classes   of nontrivial 
superconformal theories within the  context
of   $N=2$  supersymmetric theories \cite{SCFT}.   They  occur  at some special points of QMS (space of vacua)   
and/or for particular value of the parameters (such as bare masses),   for  different gauge
groups and with different matter contents,  but   fall into  various  universality classes. 
They have been classified by    Eguchi et. al.\cite{Eguchi}

\section{New mechanisms of confinement/dynamical symmetry breaking   in    $SU(n_c), USp(2n_c) $ and  $SO(n_c)$  gauge theories  }

Recently  the  questions   such as:    
{i)} the mechanism of confinement; 
{ii)} the mechanism of flavor (chiral) symmetry breaking;
  and the relation between the two; 
{iii)} the existence of other phases (CFT, oblique confinement, etc.), 
have been  studied in detail  \cite{CKM,CKKM}  in a large   class of models based on $SU(n_c), USp(2n_c) $ and  $SO(n_c)$ 
gauge groups  and with $N=2$ supersymmetry,  where   the    adjoint scalar  mass term breaks  the supersymmetry   
to $N=1$.

 The most striking results  of our analysis,  
 summarized in Table 3  and  Table 4  for $SU(n_c)$  and  $USp(2n_c)$    theories, 
 are the  following.   
 
\begin{table}[h]
\begin{center}
\vskip .3cm
\begin{tabular}{|l|c|c|c|} 
\hline    
    Deg.Freed.      &  Eff. Gauge  Group
&   Phase    &   Global Symmetry     \\
\hline \hline
  monopoles   &   $U(1)^{n_c-1} $               &   Confinement
   &      $U(n_f) $            \\ \hline
  monopoles         & $U(1)^{n_c-1} $        &
Confinement       &     $U(n_f-1) \times  U(1) $        \\ \hline
 dual quarks        &    $SU(r)
\times U(1)^{n_c-r}   $  &    Confinement
&          $U(n_f-r) \times U(r) $
\\ \hline
  rel.  nonloc.     &    -    &    Almost SCFT
&          $U({n_f / 2} ) \times U({n_f/2}) $
\\ \hline
  dual quarks     &
$ SU({\tilde n}_c) \times  U(1)^{n_c -  {\tilde n}_c } $                &  Free Magnetic
&      $U(n_f) $         \\ \hline
\end{tabular}
\caption{  Phases of $SU(n_c)$ gauge theory with $n_f$ flavors.  The label $r$ in the third row runs 
for $r=2,3,\ldots,  [{n_f -1\over 2}].\,   $   ``rel. 
nonloc."  means that  
 relatively nonlocal monopoles and dyons   coexist  as low-energy effective degrees of freedom.   
``Confinement" and ``Free Magnetic" refer to phases with $\mu \neq 0$.
``Almost SCFT" means
that the theory is a non-trivial superconformal one for $\mu=0$ but confines with $\mu \neq 0$.  
 $ {\tilde n}_c \equiv n_f-n_c.$  }
\end{center}
\label{tabsun1}  
\end{table}     
 
\begin{table}[h]
\begin{center}
\vskip .3cm
\begin{tabular}{|l|c|c|c|}
\hline
  Deg.Freed.      &  Eff. Gauge Group
&   Phase    &   Global Symmetry     \\
\hline \hline 
  rel.  nonloc.       &    -    &
Almost SCFT
&          $ U(n_f)  $
\\ \hline
 dual quarks     &      $USp(2  {\tilde n}_c) \times  U(1)^{n_c -{\tilde n}_c} $          
     &  Free
Magnetic &      $SO(2n_f) $         \\ \hline
\end{tabular}
\caption{ Phases of $USp(2 n_c)$ gauge theory  with $n_f$ flavors  with $m_i \to 0$.  
 $ {\tilde n}_c \equiv n_f-n_c-2$. }
\label{tabuspn}
\end{center}
\end{table}

The  't Hooft - Mandelstam 
picture of confinement,  caused by     the  condensation of 
  monopoles in the maximal Abelian subgroup $U(1)^k$,  ($k=$ Rank of the gauge group),   is in fact    realized
  but  in some of the vacua.  In a more
``typical" vacuum of $SU(n_c)$   gauge theory,    the  effective,  infrared degrees of freedom involve 
are   a set of dual  quarks, interacting with      low-energy  effective non-Abelian $SU(r)$   gauge
fields.  The 
condensation of these magnetic quarks as well as  of certain Abelian monopoles also present in the theory,   upon $\mu$
perturbation, lead to   confinement and dynamical symmetry breaking.  The
semi-classical monopoles may be interpreted as baryonic composites made of these magnetic quarks
and monopoles,    which  break up into
their constituents before they become massless,  as  we move from 
the semiclassical region of   the space of  $N=2$ vacua (parametrized by a set of gauge invariant VEVS)
towards the relevant singularity.    These theories are essentially infrared-free.

The second most interesting result is that    the special vacua  ($r=n_f/2$ in Table 3)   in
$SU(n_c)$ theory as well as the entire first group of vacua in $USp(2n_c)$ or   $SO(n_c)$      theory,      correspond to
various   nontrivial infrared fixed points (SCFT).    The low-energy effective degrees of freedom in general contain
relatively nonlocal  states and   there is no local effective  Lagrangian description of these theories,  though the symmetry
breaking pattern can be found from the analysis at large adjoint mass $\mu$.    The symmetry breaking pattern 
of these vacua is  $SU(n_f) \to SU(n_f/2)$,    $USp(2n_f) \to U(n_f)$  and  $SO(2n_f) \to U(n_f)$ in $SU(n_c), USp(2n_c) $ and 
$SO(n_c)$  gauge theories, respectively.

 Finally, in both type of gauge  theories,   for large number
of flavor,  there is a second group of vacua in free-magnetic phase,    with no confinement and
no  spontaneous flavor symmetry breaking.   In these vacua the low energy degrees of freedom are 
weakly interacting non-Abelian dual quarks and gauge particles,   as well as some  monopoles of products of 
$U(1)$ groups.   In $SO(n_c)$ 
theories, the situation is   qualitatively    similar \cite{CKKM}  to   $USp(2n_c)$  cases;  however,   the effective  gauge group
and  the unbroken global  group in the vacua in free-magnetic phase   are  given by
$ SO({\tilde n}_c)=SO(2n_f-n_c+4)$ and   $USp(2n_f)$, respectively, in   these   theories.

To summarise,   the  picture of confinement due to condensation of monopoles in the maximally
Abelian gauge subgroup (\`a la 't Hooft-Mandelstam)    is realized   in few of the vacua   only.     Two other mechanisms of
confinement/dynamical symmetry breaking  have been discovered.  One is the condensation of dual quarks with effective
(dual) non-Abelian gauge interactions;  the other   is   based on nontrivial SCFT (condensing entities  involve relatively
nonlocal   dyons).   In both these cases,   the maximally Abelian monopoles of $U(1)^R$ theory ($R$= rank of the gauge
group)  do not represent the correct low-energy degrees of freedom.

\section*{Acknowledgments}

K.K.   thanks  G. Carlino,  P.Kumar,   H. Murayama for  enjoyable and fruitful  collaboration,   F. Fucito,  K. Fujikawa, 
Tim Morris, N. Sakai  and G. Veneziano for useful discussions at various stages.   Special thanks to the organizers of the
conference,  S. Arnone,  Tim Morris and K. Yoshida  for  providing  him  with  the occasion to  discuss  some of  the latest
results in a stimulating atmosphere.  Part of the work was  done while the author was visiting the Theory Division of CERN,
whose hospitalty is warmly acknowledged.

\end{document}